\documentclass[article,moreauthors,pdftex,10pt,a4paper]{mdpi} 

\firstpage{1} 

\pdfoutput=1

\usepackage{bm}
\renewcommand{\v}[1]{\boldsymbol{#1}}
\renewcommand{\c}[1]{\mathcal{#1}}
\newcommand{\m}[1]{\mathrm{#1}}

\Title{PHENIX results on three-particle Bose-Einstein correlations
in $\sqrt{S_{NN}} = 200$ GeV Au+Au collisions}

\Author{Tam\'as Nov\'ak for the PHENIX Collaboration}

\AuthorNames{Tam\'as Nov\'ak}
\address[1]{
Eszterh\'ay K\'aroly University K\'aroly R\'obert Campus, H-3200 Gy\"ongy\"os, M\'atrai \'ut 36, Hungary
}
\corres{Correspondence: novak.tamas@uni-eszterhazy.hu; Tel.: +36-37-518-300}

\abstract{Bose-Einstein correlations of identical hadrons reveal information about hadron creation from the strongly interacting matter formed in ultrarelativistic heavy ion collisions. The measurement of three-particle correlations may in particular shed light on hadron creation mechanisms beyond thermal/chaotic emission. In this paper we show the status of PHENIX measurements of three pion correlations as a function of momentum differences within the triplets. We analyze the shape of the correlation functions through the assumption of L\'evy sources and a proper treatment of the Coulomb interaction within the triplets. We measure the three-particle correlation strength ($\lambda_3$), which, together with the two-particle correlation strength $\lambda_2$, encodes information about hadron creation mechanisms. From a consistent analysis of two- and three-particle correlation strength we establish a new experimental measure of thermalization and coherence in the source.
}

\keyword{RHIC, PHENIX, Bose-Einstein correlations, L\'evy distribution, thermalization, coherence}

\conferencetitle{10th Bolyai-Gauss-Lobachevsky Conference on Non-Euclidean Geometry and its Applications}

\usepackage{subcaption}
\usepackage{chngcntr}
\preto{\abstractkeywords}{\nolinenumbers}

\begin{document}

\section{Introduction}

In particle and nuclear physics, intensity interferometry provides a direct experimental method
for the determination of sizes, shapes and lifetimes of particle-emitting sources (for reviews
see \citep{gyulassy, boal, baym, kittel, csorgo_2002}). In particular, boson interferometry provides a powerful tool for the investigation
of the space-time structure of particle production processes, since Bose-Einstein correlations
(BEC) of two or three identical bosons reflect both geometrical and dynamical properties of the particle-radiating source.

The size (radius) of the source in heavy-ion collisions has been found to decrease with
increasing transverse momentum, $p_t$, or transverse mass, $m_t = \sqrt{m^2 + p_t^2}$, of the bosons. This effect can also be explained by hydrodynamical models \citep{heinz, csorgo_1996}. The main purpose of the present paper is to determine the three particle correlation strength ($\lambda_3$) as function of transverse momentum. 
Recently \citep{two-pion}, the two-particle correlation strength ($\lambda_2$) was determined. Those measurements of $\lambda_2$, when combined with this analysis, may test the limits of Core-Halo model \citep{Bolz:1992hc, Csorgo:1994in, csanad_2005} with a thermalized core. For this purpose we introduce a new parameter, $\kappa_3$, which is function of $\lambda_3$ and $\lambda_2$. This new parameter is not equal with $1$ when there are extra effects in the core, for example not fully thermalized core, or partial coherence in the core. The main purpose of the present work is to investigate whether $\kappa_3$ indicates extra effects or not.

Let us start with the definition of the three-particle correlation function:

\begin{equation}
C_3(\bm{k_1}, \bm{k_2}, \bm{k_3})=\frac{N_3(\bm{k_1}, \bm{k_2}, \bm{k_3})}{N_1(\bm{k_1})N_1(\bm{k_2})N_1(\bm{k_3})},
\end{equation}
where $N_3$ is the three particle invariant momentum distribution, defined by

\begin{equation}
N_3(\bm{k_1}, \bm{k_2}, \bm{k_3}) = \int \c S(\bm{r_1}, \bm{k_1}) \c S(\bm{r_2}, \bm{k_2})\c S(\bm{r_3}, \bm{k_3})|\Psi_{\bm{k_1, k_2, k_3}}(\bm{r_1},\bm{r_2},\bm{r_3})|^2 \Pi_{i=0}^{3} d^4 \bm{r_i}
\end{equation}
and $N_1$ is the single particle invariant momentum distribution, defined as

\begin{equation}
N_1(\v k) = \int \c S(\v r_1, \v k_1) |\Psi_{\v k}(\v r_1)|^2 d^4 \v r.
\end{equation}
In the above equations, $\Psi_{\v k_1, \v k_2, \v k_3}$ is the three-particle wave function, and $\Psi_{\v k}$ is the single particle wave function. Furthemore, $S(\v r, \v k)$ is the source distribution which describes the probability density of particle creation at the space-time point $\v r$ with momentum $\v k$.

Our assumption for the source function is the symmetric L\'evy distribution \citep{nolan, metzler, csorgo_2004, csanad, csanad_2017}, which is defined by:

\begin{align}\label{e:L\'evydef}
L(\alpha,R,\v r)=\frac{1}{(2\pi)^3}\int\m d^3\v q\, e^{i\v q\v r} e^{-\frac{1}{2}|\v qR|^{\alpha}} ,
\end{align}
where $\alpha$ is the L\'evy index and $R$ is the L\'evy scale. Then $\alpha=2$ gives back the Gaussian case and $\alpha=1$ yields a Cauchy distribution. For the L\'evy stable source distribution, 0 < $\alpha \le 2$ in general.

Assuming properly symmetrized plane-waves for the wave functions, and the L\'evy distribution as source function, the three particle correlation function is given as

\begin{align}
C_3^{(0)}(k_{12}, k_{13}, k_{23}) = 1+ \ell_3e^{-0.5(|2k_{12}R|^\alpha+|2k_{13}R|^\alpha+|2k_{23}R|^\alpha)}
+\ell_2\bigg(e^{|2k_{12}R|^\alpha}+e^{|2k_{13}R|^\alpha}+e^{|2k_{23}R|^\alpha}\bigg),
\end{align}
where $k_{ij}=|{\v k}_i - {\v k}_j|/2$, i.e. the half momentum difference of the $i$th and $j$th particle. Furthermore, the $\ell_2$ parameter is the two particle correlation strength parameter in the three particle correlation function, and the $\ell_3$ is the three particle strength parameter.

To be able to fit the model to the data, the correlation function has to be completed with a background which parametrizes possible long-range correlations. In this analysis we used a simple linear background in every direction, with the same slope. Then the correlation function is the following:

\begin{equation}
C_{3, \mathrm{fit}}^{(0)}(k_{12}, k_{13}, k_{23})= N(1+\epsilon k_{12})(1+\epsilon k_{13})(1+\epsilon k_{23})C_3^{(0)}(k_{12}, k_{13}, k_{23}).
\end{equation}

We also have to take into account the Coulomb interaction since this decreases the number of particle pairs at low momentum differences. To add Coulomb correction to the model, we define a Coulomb correction factor:

\begin{equation}
K_3(\boldsymbol{k_{12}},\boldsymbol{k_{23}},\boldsymbol{k_{31}})=\frac{\int{d^3\boldsymbol{r_1}}d^3\boldsymbol{r_2}d^3\boldsymbol{r_3}\c S(\boldsymbol{r_1})\c S(\boldsymbol{r_2})\c S(\boldsymbol{r_3})\vert\Psi_{\boldsymbol{k_{12}},\boldsymbol{k_{23}},\boldsymbol{k_{31}}}^0(\boldsymbol{r_1},\boldsymbol{r_2},\boldsymbol{r_3})\vert^2}{\int{d^3\boldsymbol{r_1}}d^3\boldsymbol{r_2}d^3\boldsymbol{r_3}\c S(\boldsymbol{r_1})\c S(\boldsymbol{r_2})\c S(\boldsymbol{r_3})\vert\Psi_{\boldsymbol{k_{12}},\boldsymbol{k_{23}},\boldsymbol{k_{31}}}^C(\boldsymbol{r_1},\boldsymbol{r_2},\boldsymbol{r_3})\vert^2},
\end{equation}
where the $\Psi^0$ is the free particle wave function (symmetrized plane wave), and the $\Psi^C$ is the (symmetrized) solution of full three particle Coulomb problem. Instead of solving the three particle Coulomb problem, we used a general approach to handle the three particle Coulomb interaction, which is called ''Generalized Riverside'' method~\citep{dgangadharan}:

\begin{equation}
K_3(k_{12}, k_{13}, k_{23}) \approx K_2(k_{12})K_2(k_{13})K_2(k_{23}),\;\;\textnormal{ with }\;\;
K_2(k_{ij})=\frac{
\int{d^3\boldsymbol{r_1}}d^3\boldsymbol{r_2}\c S(\boldsymbol{r_1})\c S(\boldsymbol{r_2})
\vert\Psi_{\boldsymbol{k_{12}}}^0(\boldsymbol{r_1},\boldsymbol{r_2})\vert^2}{
\int{d^3\boldsymbol{r_1}}d^3\boldsymbol{r_2}\c S(\boldsymbol{r_1})\c S(\boldsymbol{r_2})
\vert\Psi_{\boldsymbol{k_{12}}}^C(\boldsymbol{r_1},\boldsymbol{r_2})\vert^2}.
\end{equation}
This approximates the three particle Coulomb correction well if certain conditions \citep{dgangadharan} are satisfied, and does not depend on the directions of the momentum
differences, just their magnitude. With this Coulomb correction factor, the full Bose-Einstein correlation function is

\begin{equation} 
C_3(k_{12}, k_{13}, k_{23}) = C_{3, \mathrm{fit}}^{(0)}(k_{12}, k_{13}, k_{23}) K_{Coulomb}(k_{12}, k_{13}, k_{23}).
\end{equation}

By using this approximation we are able to reuse the Coulomb correction calculation method, developed in \citep{Alt:1998nr}. After some consistency checks we conclude that using this approximation can be considered since the relative systematic error due to this approximation is less than 2\%. 

The two- and three-particle correlation strengths are defined as the extrapolated intercept of the given correlation function:

\begin{align}
\lambda_2 &\equiv C_2(k_{12}\rightarrow 0)-1,\\
\lambda_3 &\equiv C_3(k_{12}=k_{13}=k_{23}\rightarrow 0)-1.
\end{align}  
With respect to these, it is important to note that not all pions are created directly from the strongly interacting matter. A significant fraction of pions are secondary, coming from decays. Hence the source will have two components: a core of primordial pions, and a halo consisting of the decay products of long lived resonances (such as $\eta$, $\eta^\prime$, $K_s^0$, $\omega$)

\begin{equation}
S = S_{core} + S_{halo}.
\end{equation}  
In the Core-Halo model, the fraction of the core is defined by:

\begin{equation}
f_c=\frac{N_\mathrm{core}}{N_\mathrm{core}+N_\mathrm{halo}},
\end{equation}
where the $N_\mathrm{core}$ is the number of pions in core, and $N_\mathrm{halo}$ is the number of pions in the halo.
If there is partial coherence in core, a coherence parameter can be introduced~\cite{Csorgo:1998tn, NA44}, with examining the fraction of coherently produced pions in core:

\begin{equation}
p_c=\frac{N_\mathrm{coherent}}{N_\mathrm{coherent}+N_\mathrm{incoherent}},
\end{equation}
where the $N_\mathrm{coherent}+N_\mathrm{incoherent}=N_\mathrm{core}$.
With these, $\lambda_2$ and $\lambda_3$ can be expressed as functions of $f_c$ and $p_c$:

\begin{align}
\lambda_2=& f_C^2\big[(1-p_c)^2+2p_c(1-p_c)\big],\label{e:pc1}\\
\lambda_3=& 2f_C^3\big[(1-p_c)^3+3p_c(1-p_c)^2\big]+3f_C^2\big[(1-p_c)^2+2p_c(1-p_c)\big].\label{e:pc2}
\end{align}

To be able to investigate the partial coherence within the limits of this model, we have to take Equations (\ref{e:pc1})--(\ref{e:pc2}) and use the measured $\lambda_2$ and $\lambda_3$ parameters as input in these equations, and we have to look for $f_c$, $p_c$ values which solves both equation. To determine whether a solution exists, we can plot $f_c$ as function of $p_c$ with given $\lambda_2$ and $\lambda_3$. In this way we get two curves (one from equation with $\lambda_2$, an other from equation with $\lambda_3$), and if these curves overlap then the second and the third order Bose-Einstein correlation functions are consistent with each other and also the degree of partial coherence can be determined from region of the overlap.

\section{Results}
\begin{figure}
\centering
\includegraphics[width=0.49\linewidth]{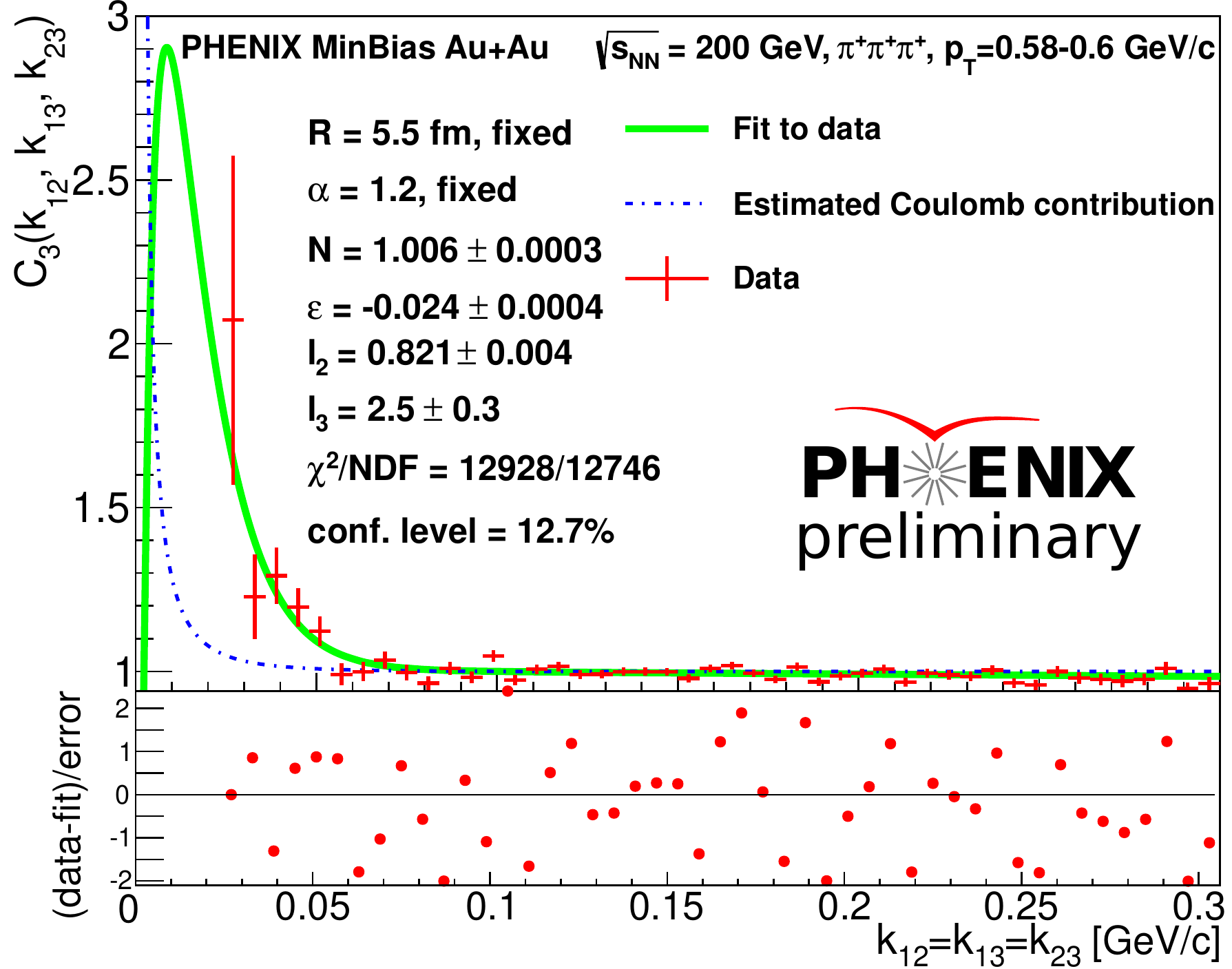}
\includegraphics[width=0.49\linewidth]{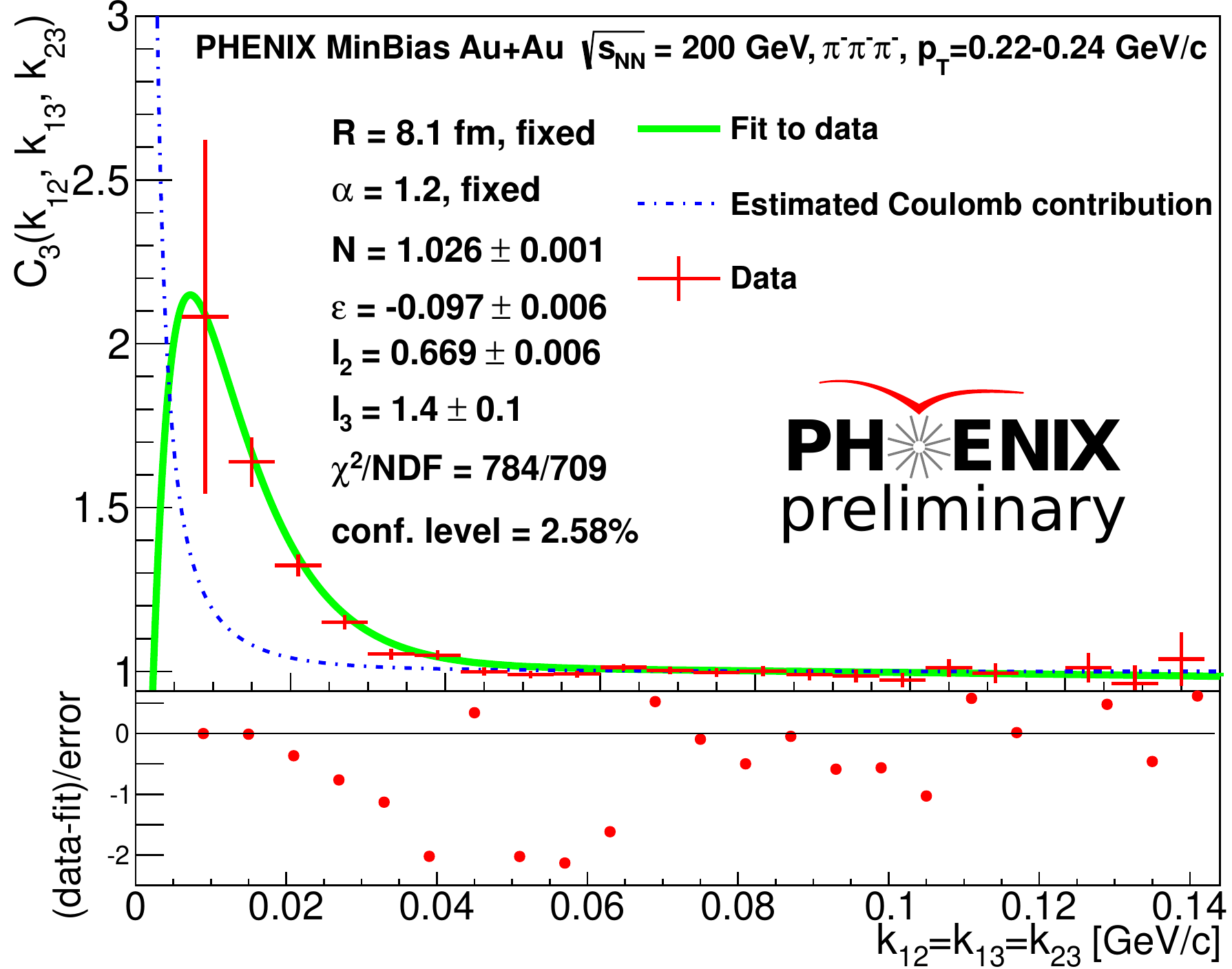}
\caption{Diagonal visualization of the three-particle Bose-Einstein correlation functions $C_3$ with the results of fits.
\label{f:fits}} 
\end{figure}

The data used in the analyses were collected by the PHENIX detector in 2010. 7.3 billion minimum bias Au+Au collisions at $\sqrt{s_{NN}} = 200$ GeV were recorded. We measured three-particle correlation fnctions of $\pi^- \pi^- \pi^-$ and $\pi^+ \pi^+ \pi^+$ triplets in 31 $m_T$ bins ranging from 228 to 871 MeV, with event selection, single track cuts and pair cuts 
identical to the analysis of Ref. \citep{two-pion}. The visualization of three-dimensional fits is difficult, so instead of showing two-dimensional projections of these three-dimensional fits, we decided to show, as an illustration, various one plus one dimensional slices of them. These slices were obtained by requiring that the two-particle relative momenta $k_{i,j}$ between the pairs $(i,j) = (1,2)$ , $(2,3)$ and $(3,1)$ are equal.
The $\chi^2/NDF$ and the confidence levels are statistically acceptable. 
Figure~\ref{f:fits} indicates the PHENIX preliminary three-particle correlation data and the fitted curve as a univariate slice of a multi-variate function. Note that no significant long-range correlation is observed: $\epsilon$ is close to zero. 
In these fits, the values of the L\'evy-scale $R$ and the L\'evy index of stability $\alpha$ were taken from PHENIX preliminary two-particle correlation data in the same transverse momentum bins, and in the same selection of minimum-bias, $\sqrt{s_{NN}} = 200$ GeV Au+Au collisions. See these preliminary $R$ and $\alpha$ data in Ref.~\cite{Kincses:2016jsr}. Within one standard deviation the same parameter values are obtained if we release the values of $R$ and $\alpha$ in these fits, indicating an internal consistency between the two-particle and the three-particle correlation measurements. Given this internal consistency and  our focus on the comparison of the strength or intercept parameters $\lambda_2$ and $\lambda_3$ of the two and the three-particle Bose-Einstein correlation functions,  here we report these values using the $R$ and $\alpha$ values as determined from two-particle Bose-Einstein correlation function measurements in miminum bias $\sqrt{s_{NN}} = 200$ GeV Au+Au collisions~\citep{Kincses:2016jsr}.

\subsection{Three-particle correlation strength}

\begin{figure}
\centering
\includegraphics[width=0.99\linewidth]{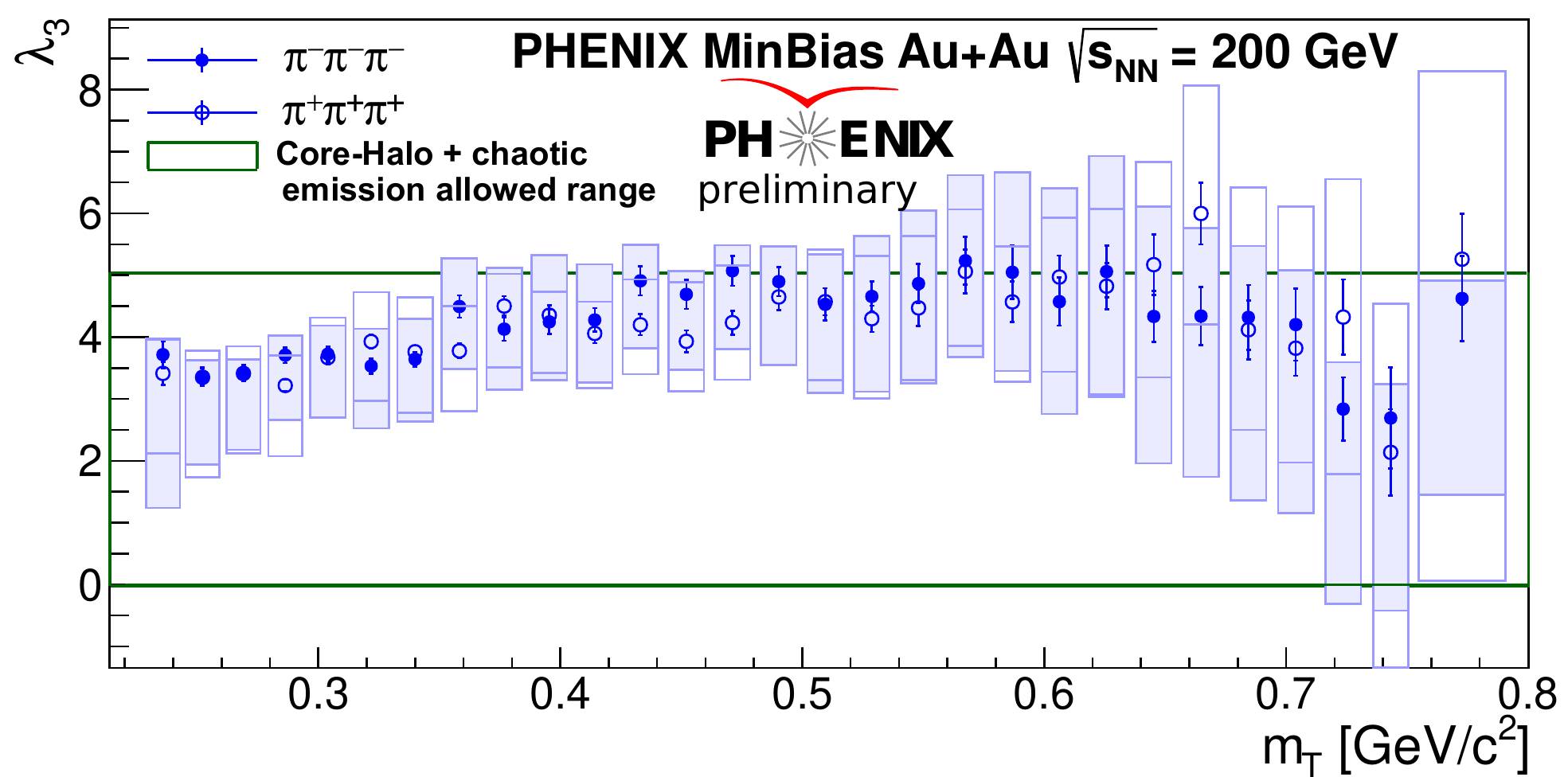}
\caption{The three-particle correlation strength, $\lambda_3$, as function of $m_T$. Statistical and systematic uncertainties shown as bars and boxes, respectively.} 
\label{fig:lambda_3}
\end{figure}

The three-particle correlation strength, derived from $\ell_2$ and $\ell_3$ fit parameters as $\lambda_3=\ell_2+\ell_3$, can be seen in Figure~\ref{fig:lambda_3}. According to the Core-Halo model, $\lambda_3$ has to be in the range of $[0, 5]$ indicated by a green box in Figure~\ref{fig:lambda_3}. One can see that within errors this parameter falls in the allowed range. The detailed description of the systematic uncertainty sources is given in Ref. \citep{two-pion}.

\subsection{Core-Halo independent parameter}

In this subsection the possibility of partial coherence is investigated. We introduce a new parameter, a combination of $\lambda_3$ and $\lambda_2$ as follows:

\begin{equation}
\kappa_3 = \frac{\lambda_3-3\lambda_2}{2\sqrt{\lambda_2^3}}.
\end{equation}

This parameter does not depend on value $f_c$, and its value is unity, independently of the transverse mass, if the particle emission has no coherent component, or if $p_c = 0$ in the Core-Halo picture of Bose-Einstein $n$-particle correlations.

\begin{figure}
\centering
\includegraphics[width=0.99\linewidth]{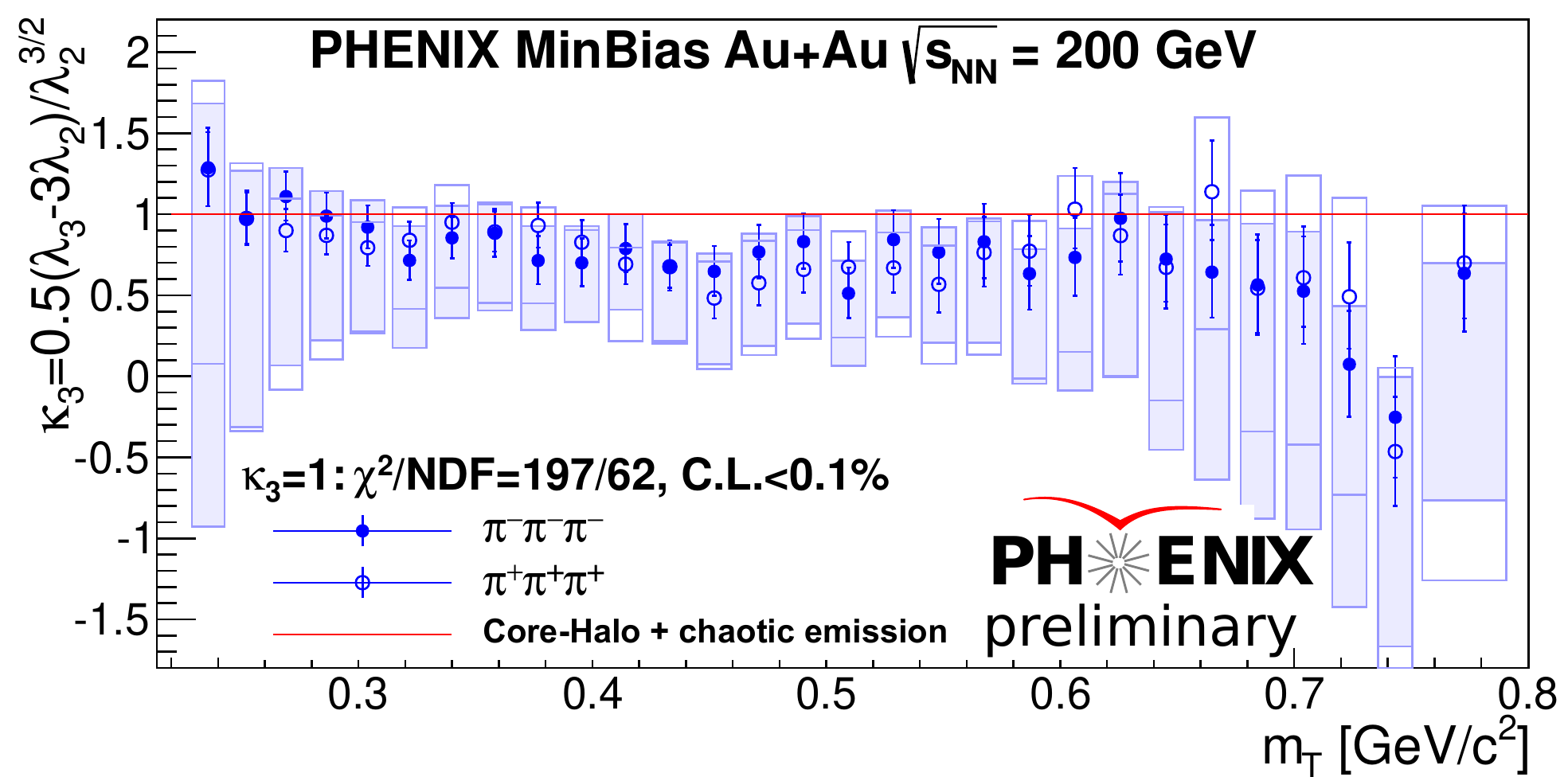}
\caption{\label{f:kappa3}
Core-Halo independent parameter as function of $m_T$. Statistical and preliminary systematic uncertainties shown as bars and boxes, respectively.}
\end{figure}

The measured $\kappa_3$ parameter is shown in Figure~\ref{f:kappa3}, calculated by using $\lambda_3$ from the fits described in this paper, and $\lambda_2$ from Ref. \citep{Kincses:2016jsr}. Our errors are still preliminary, a full systematic analysis is in progress. Figure~\ref{f:kappa3} shows the $\kappa_3$ parameter for average $m_T$ of the triplets.

\begin{figure}
\centering
\includegraphics[width=0.495\textwidth]{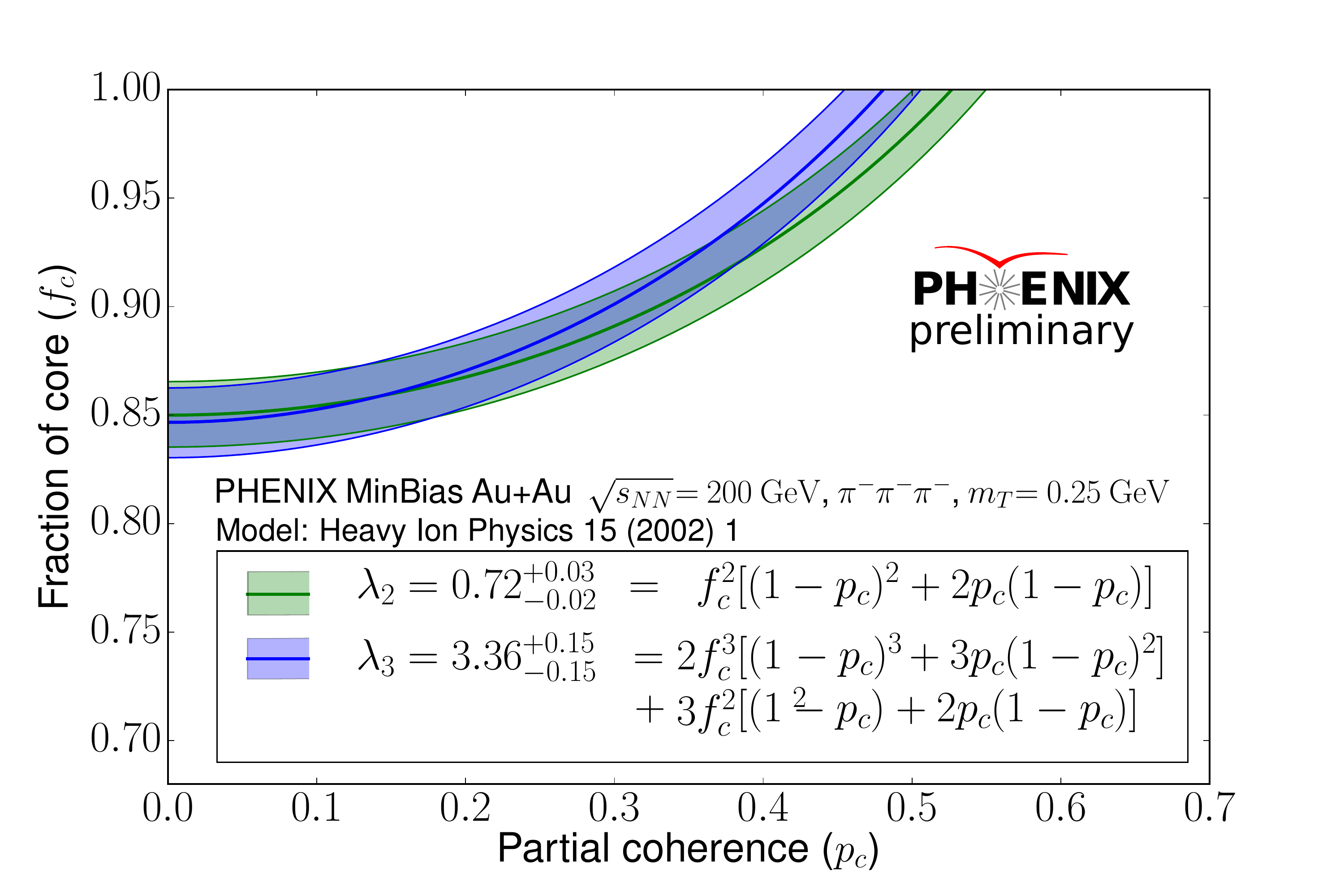}
\includegraphics[width=0.495\textwidth]{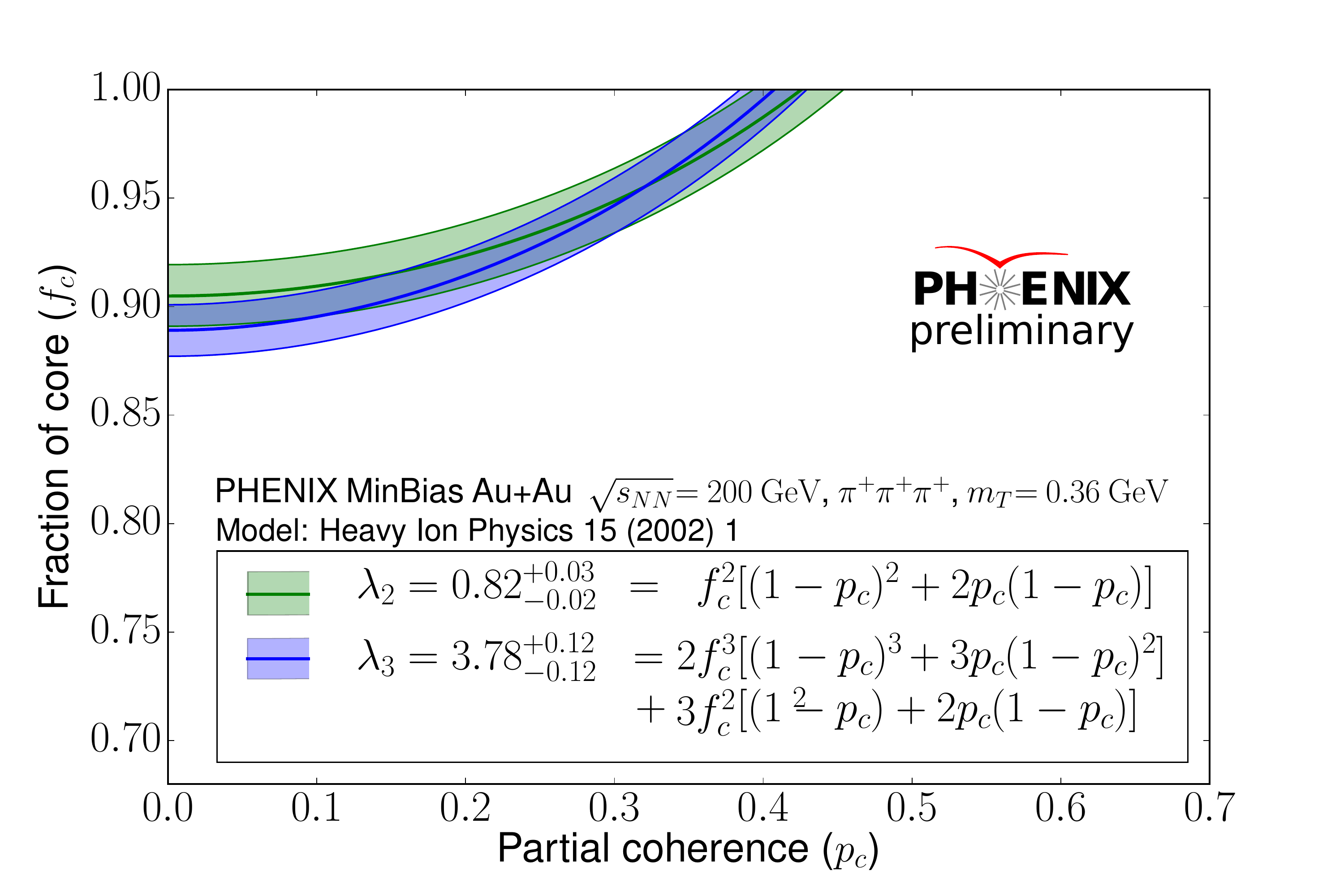}
\vspace{-10pt}
\caption{\label{f:fcpc}
The fraction of the core $f_c$ parameter as a function of coherence parameter $p_c$ for two different $m_T$ values.} 
\end{figure}

We observe that $\kappa_3$ is different from unity at $m_T \approx 450$ MeV. 
The deviation of $\kappa_3$ from one may indicate some kind of partial coherence, however, the preliminary status of these datapoints and their error bars prevents us from a more definitive conclusion, further investigations are on-going and will be reported elsewhere. Figure~\ref{f:fcpc} shows $f_c$ as a function of $p_c$ for different $m_T$ regions. Here we took Equations (\ref{e:pc1})--(\ref{e:pc2}) and the measured $\lambda_2$ and $\lambda_3$ parameters, and determined the allowed $(f_c,p_c)$ region,
as shown in Figure~\ref{f:fcpc}.

\section{Conclusions}
In this paper, we simultaneously investigated two- and three-particle femtoscopic correlation functions in PHENIX Au+Au collisions at 200 GeV/nucleon collision energy. Lévy distributions yield correlation functions which provide a statistically acceptable description of the measurements. In this paper, we used source parameters $R$ and $\alpha$ from two-particle fits, and obtained the three-particle correlation strength from the analysis of three-particle correlations. We investigated the results in view of the core-halo model, and derived a core-halo independent parameter $\kappa_3$. These preliminary results were compared to predictions based on chaotic emission, and very small deviations were observed. This however needs a detailed final analysis, which we plan to carry out in the near future.

\acknowledgments{T. N. was supported by the EFOP-3.6.1-16-2016-00001 grant. This work was also supported by NKFIH grants FK-123942 and FK-123959.}

\end{document}